\documentclass[aps,prc,preprint,groupedaddress]{revtex4-1}

\begin{document}


\title{Faddeev-type equations for three-body symmetry violating scattering amplitudes}


\author{Vladimir Gudkov}
\email[]{gudkov@sc.edu}

\author{Young-Ho Song}
\email[]{song25@mailbox.sc.edu}
\affiliation{Department of Physics and Astronomy, University of South Carolina, Columbia, SC, 29208}


\date{\today}

\begin{abstract}
The equations which relate three-body and two-body symmetry violating scattering amplitudes are derived
in the first order of symmetry violating interactions.
They can be used to obtain three-body symmetry violating scattering amplitudes from two-body symmetry violating scattering amplitudes
 calculated in low energy effective field theory.

%
%
\end{abstract}


\maketitle


The study of parity violating (PV) and time reversal invariance violating (TRIV) effects in low energy physics are very important problems for understanding  main features of the Standard model and for a search for new physics.
During the past 50 years many calculations of
different PV and TRIV effects in nuclear physics have been done.
However, in the last few years it became clear (see, for example \cite{Zhu:2004vw,HolsteinUSC,DesplanqueUSC,RamseyMusolf:2006dz}  and references therein) that the traditional DDH \cite{Desplanques1980} method for the calculation of PV effects cannot reliably describe the available experimental data.
It could be blamed on the ``wrong'' experimental data,
however, it may be that DDH approach is not adequate for the description of the set of precise experimental data because it is based on a number of models and assumptions.
Recently a new approach based on the effective field theory (EFT) has been introduced as a model independent parameterization of PV effects (see, papers \cite{Zhu:2004vw,RamseyMusolf:2006dz} and references therein), and some calculations for two-body systems have been done \cite{Liu:2006dm}.
The power of the EFT approach could be utilized if we can analyze a large enough number of PV effects to be able to constrain all free parameters of the theory, which usually called as low energy constants (LEC), to guarantee the adequate description of the strong interaction hadronic part of symmetry violating observables.
Then, if discrepancies between experimental data and EFT calculations will persist, it will be a clear indication that the problems are related to weak interactions in nuclei and probably to a manifestation of new physics.

Unfortunately, the number of experimentally measured (and independent in terms of unknown constants) PV effects in two body systems is not enough to constrain all LECs. In spite of the fact, that five independent observable parameters in two body system would fix  five unknown PV LECs \cite{Girlanda:2008ts,Phillips:2008hn,Shin:2009hi,Schindler:2009wd},
 it is impossible to measure all of them using existing experimental techniques.
Therefore, one has to include into analysis few-body systems and even heavier nuclei, which are actually preferable from experimental point of view, because usually the measured effects in nuclei are much larger than in nucleon-nucleon scattering due to nuclear enhancement factors \cite{Sushkov:1982fa,Bunakov:1982is,Gudkov:1991qg}.
To verify the applicability of the EFT  approach for calculations of symmetry violating effects in  nuclear reactions, it is natural to start from a scattering problem in  three-body systems, and to develop a regular and self consistent approach for calculation of symmetry violating amplitudes in a few-body systems, which later could be extended to many body systems.

 Since symmetry violating effects are usually very small (especially for PV and TRIV), the distorted wave Born approximation (DWBA) approach is a standard and an efficient method to calculate symmetry violating amplitudes with a very good accuracy. Calculations of scattering amplitudes for three-body systems could also be done using DWBA with three-body wave-functions obtained from a solution of Faddeev equations (see, for example, recent calculation for PV neutron spin rotation in neutron-deuteron scattering \cite{Schiavilla:2008ic}).
However, this method, providing numerical values for symmetry violating observables, does not relate symmetry violating three-body and two-body amplitudes (see appendix), which is a crucial condition for a systematic extension of two-body EFT formalism  on many-body systems. To relate three-body symmetry violating scattering amplitudes directly to  symmetry violating amplitudes obtained in the EFT for two-body process, we propose a new form of Faddeev equations.

We describe a three-body system by the Hamiltonian  with  interactions $V_{ij}$ between $i$ and $j$ particles as
\begin{equation}\label{h}
H=H_0+\sum_{\gamma}V_{\gamma}; \; V_{\gamma}=V_{ij} \; \text{if} \; \gamma \neq i,j; \; \; \gamma , i,j = 1,2,3,
\end{equation}
where $H_0$ is the operator of kinetic energy
and $V_{ij}$ includes both symmetry conserving (strong) interactions and
symmetry violating interactions. For the sake of simplicity we do not include three-body forces. Should it be necessary, three-body forces could be included in a  transparent way  and will not change the final result. However,   it was shown \cite{Zhu:2004vw} that  for  PV effects, there is no contribution in a leading order of the EFT from three-body forces, and therefore, we will not consider it here. The same arguments can be applied for time reversal violating interactions because of the similar structure of these interactions on the level of nucleon-pion degrees of freedom. Therefore, one can expect the same suppression of three-body time reversal violating effects (unless we are  considering exotic parity conserving time reversal violating interactions).

Then, the scattering process can be described (see, for example \cite{FB} and references therein) in terms of transition amplitudes $<\phi_\beta |U_{\beta \alpha}|\phi_\alpha> \equiv <\phi_\beta |V^{\beta }|\Psi^{(+)}_\alpha>$ from a channel $\alpha$ to a channel $\beta$, where $\Psi^{(+)}_\alpha$ is the wave function for the scattering state of an initial channel, $\phi_\alpha$ and $\phi_\beta$ are wave functions of initial and final states, and $V^{\beta }\equiv V_{\gamma }+V_{\delta }$ (with $\beta\neq\gamma\neq\delta$) is the interaction between the particles in channel $\beta$.

Since we are interested in the scattering problem, we will use Faddeev equations \cite{Faddeev:1960su} written in terms of transition operators $U_{\beta \alpha}$ and known as the AGS-equations \cite{AGS}:
\begin{equation}\label{AGS}
    U_{\beta \alpha}=\bar{\delta}_{\alpha \beta}G^{-1}_0 + \sum_{\gamma}\bar{\delta}_{\gamma \beta}t_{\gamma}G_{0}U_{\gamma \alpha},
\end{equation}
where $\bar{\delta}_{\alpha \beta}=1-\delta_{\alpha \beta}$,  $t_{\gamma}$ is a two-particle transition operator in three-particle space, and
\begin{equation}\label{freeres}
   G_0(z)=\frac{1}{z-H_0}
\end{equation}
is the resolvent operator for free motion.

Let us analyze these equations for the case of two types of interactions between the particles: one of which (a ``regular'' one) is  preserving the symmetry and another one is violating the symmetry, and assume that the symmetry violating interactions are much smaller than  regular ones. As an example, one can consider parity conserving (PC) and parity violating (PV)interactions. In this, case both  two-particle and tree-particle scattering operators can be represented as a sum of PC (indicated by $s$ for ``strong'') and PV (indicated by $w$ for ``weak'') parts:
\begin{equation}\label{texp}
   t_\gamma=t^s_\gamma +t^w_\gamma
\end{equation}
and
\begin{equation}\label{Uexp}
   U_{\beta \alpha}=U^s_{\beta \alpha}+U^w_{\beta \alpha},
\end{equation}
which satisfy the following inequalities $|<\ldots |t^s_\gamma |\ldots >| \gg |<\ldots |t^w_\gamma |\ldots >|$ and $|<\ldots |U^s_{\beta \alpha} |\ldots >| \gg |<\ldots |U^w_{\beta \alpha} |\ldots >|$.
It should be noted that these two $s$- and $w$-parts are distinguished both by their values and their symmetry properties.

To obtain equations for these PC and PV operators, we substitute Eqs.(\ref{texp}) and (\ref{Uexp}) into Eq.(\ref{AGS}). Then, the resulting equations contain a sum of terms with two different symmetries: scalar and pseudo-scalar ones. To satisfy the equations, a sum of scalar and pseudo-scalar terms must be independently equal to zero , which leads to a set of coupled equations
\begin{eqnarray}
   U^s_{\beta \alpha}&=&\bar{\delta}_{\alpha \beta}G^{-1}_0 + \sum_{\gamma}\bar{\delta}_{\gamma \beta}t^s_{\gamma}G_{0}U^s_{\gamma \alpha}+\sum_{\gamma}\bar{\delta}_{\gamma \beta}t^w_{\gamma}G_{0}U^w_{\gamma \alpha}, \label{coupled1} \\
   U^w_{\beta \alpha}&=& \sum_{\gamma}\bar{\delta}_{\gamma \beta}t^s_{\gamma}G_{0}U^w_{\gamma \alpha}+\sum_{\gamma}\bar{\delta}_{\gamma \beta}t^w_{\gamma}G_{0}U^s_{\gamma \alpha},\label{coupled2}
\end{eqnarray}
where the first one preserves the symmetry and the second one violates it.
One can see that the last term in Eq.(\ref{coupled1}) has the second order in weak symmetry violating interactions, and, therefore, could be ignored since we are interested only in the first order of symmetry violating effects. Then, the above set of equations can be written as two decoupled equations: the first one (in the first order of ``weak'' interaction) for a ``strong'' symmetry conserving transition operator
\begin{equation}\label{strong}
 U^s_{\beta \alpha}=\bar{\delta}_{\alpha \beta}G^{-1}_0 + \sum_{\gamma}\bar{\delta}_{\gamma \beta}t^s_{\gamma}G_{0}U^s_{\gamma \alpha},
\end{equation}
and the second one (exact) for a ``weak'' symmetry violating operator
\begin{equation}\label{weak}
   U^w_{\beta \alpha}=
    \sum_{\gamma}\bar{\delta}_{\gamma \beta}t^s_{\gamma}G_{0}U^w_{\gamma \alpha}+\sum_{\gamma}\bar{\delta}_{\gamma \beta}t^w_{\gamma}G_{0}U^s_{\gamma \alpha}
\end{equation}
which are the main result of this paper.
As is evident the Eq.(\ref{strong}) is exactly the same as the AGS-equations (Eq.(\ref{AGS})) with interactions without symmetry violations. Therefore, a strong part of the transition operator, and, as a consequence, of the scattering amplitude can be obtained by solving the standard three-body equations.
 As concerning weak symmetry violating operator, it consists of two essentially different parts. The first term in Eq.(\ref{weak}) depends on unknown $U^w_{\beta \alpha}$ with exactly the same kernel as for strong interaction in  Eq.(\ref{strong}). The second term does not depend on $U^w_{\beta \alpha}$ and, therefore, it corresponds to a free term in three-body integral equations.
 One can see that the first (integral) term includes strong two-body transition operators but the second one (free term) contains direct contribution to $U^w_{\beta \alpha}$ from weak two-body transition operators. Therefore, Eq.(\ref{weak}) gives us a framework for a calculation of symmetry violating amplitudes using Faddeev type three-body equations in terms of two-body amplitudes, where the two-body amplitudes can be calculated using different approaches, including effective field theory \cite{Zhu:2004vw,RamseyMusolf:2006dz,Liu:2006dm}.

\begin{acknowledgments}
This work was supported by the DOE grants no. DE-FG02-09ER41621.
\end{acknowledgments}

\section{Appendix: Faddeev equations for wave function}

We can obtain equations similar to Eqs.(\ref{strong}) and (\ref{weak}) for three-body wave functions as well.
Adopting standard notations \cite{FB}, Faddeev equation for components of wave functions can be written as
\begin{equation}
(E-H_0-V_i)\psi_i=V_i(\psi_j+\psi_k)
\end{equation}
where, $i,j,k$ are cyclic permutations of $1,2,3$ representing
channels, and $V_i\equiv V(x_i)$ is a potential between
$j$ and $k$ particles, and $H_0$ is the kinetic energy of the system.
Schr\"{o}dinger wave function is given as a sum of three Faddeev
components,
\begin{equation}
\Psi=\psi_1({ x}_1,{ y}_1)+\psi_2({ x}_2,{ y}_2)+\psi_3({ x}_3,{ y}_3)
\end{equation}
where, ${ x}_i$ and ${ y}_i$ are Jacobi coordinates
describing the distance between particle $j$ and $k$ and
the distance between particle $i$ and center of cluster $j+k$, correspondingly.

For symmetry violating interaction, let us separate potential
$V_i=V_i^{s}+V_i^{w}$ and wave function $\psi_i=\psi_i^{s}+\psi_i^{w}$ as a sum of symmetry conserving and violating parts.
Then, neglecting the second order of symmetry violating contributions $V_i^w \psi_i^{w}$ and $V_i^{w} (\psi_j^{w}+\psi_k^w)$,
we have  two equations
\begin{eqnarray}
(E-H_0-V_i^s)\psi_i^s&=&V_i^s(\psi_j^s+\psi_k^s) \\
(E-H_0-V_i^s)\psi_i^w&=&V_i^s(\psi_j^w+\psi_k^w)
+V_i^w(\psi^s_i+\psi^s_j+\psi^s_k)
\end{eqnarray}
These equations have a property similar to that of derived AGS-type equations:
the first one is a Faddeev for only ``strong'' interacting particles; and the second one is equation for weak component of wave function with the same kernel as  for
``strong'' interaction.
However, these equations do not provide  transparent
relations between three-body wave function and wave functions or
 two-body scattering amplitudes.

\bibliography{AGS}

\end{document}